\documentclass[superscriptaddress,amsmath,amssymb,aps,longbibliography,twocolumn,showpacs,a4paper]{revtex4-2}
\usepackage{amsmath}
\usepackage{amsfonts}
\usepackage{multirow}
\usepackage{booktabs}
\usepackage{graphicx}
\usepackage{nicefrac}
\usepackage{subfigure}
\usepackage{xcolor}
\usepackage[normalem]{ulem}
\usepackage{bm}
\usepackage{tikz-cd}
\usepackage[breaklinks=true,pdfborder={0 0 0}]{hyperref}
\hypersetup{colorlinks=true,citecolor=red}

\makeatletter

\def\@seccntformat#1{\csname the#1\endcsname\hskip 0.5em}
\makeatother

\begin{document}

\title{Quantum benchmarks for magnetic energies and transition temperatures}
\author{Mojtaba~Alaei}
\email{m.alaei@skoltech.ru}
\affiliation{Skolkovo Institute of Science and Technology, Bolshoy Boulevard 30, bld. 1, Moscow 121205, Russia}
\affiliation{Department of Physics, Isfahan University of Technology, Isfahan 84156-83111, Iran.}
\author{Ilia~E.~Mikhailov}
\affiliation{Skolkovo Institute of Science and Technology, Bolshoy Boulevard 30, bld. 1, Moscow 121205, Russia}
\affiliation{Physical Science and Engineering Division (PSE), King Abdullah University of Science and Technology (KAUST), Thuwal, Saudi Arabia}
\author{Alireza Qaiumzadeh}
\affiliation{Center for Quantum Spintronics, Department of Physics, Norwegian University of Science and Technology, Trondheim, NO-7491, Norway}
\author{Artem~R.~Oganov}
\affiliation{Skolkovo Institute of Science and Technology, Bolshoy Boulevard 30, bld. 1, Moscow 121205, Russia}

\date{\today}
\maketitle

\section{abstract}
Predicting magnetic ordering temperatures requires accurate exchange interactions and an appropriate treatment of spin thermodynamics. Here we assess these requirements through complementary quantum Monte Carlo benchmarks. For five insulating antiferromagnets, we compare ferromagnetic--antiferromagnetic energy differences from diffusion Monte Carlo (DMC) and density functional theory with references reconstructed from inelastic neutron scattering (INS) exchanges. DMC underestimates these references but achieves a mean absolute error comparable to the best-performing functionals tested. Separately, we directly test the previously proposed $(S+1)/S$ rescaling of classical Monte Carlo temperatures against quantum stochastic series expansion simulations for 17 compounds using corresponding INS-derived Heisenberg Hamiltonians. Rescaling reduces the mean absolute classical--quantum temperature difference from 81.7 to 8.2~K, with a mean absolute percentage difference of 5.9\%. These results establish a direct quantum benchmark for the rescaling within the tested models and emphasize that improvements in electronic magnetic energy scales should be accompanied by attention to quantum spin thermodynamics.

\section{Introduction}
Magnetic ordering connects microscopic electronic interactions to a collective phase transition. Predicting the associated Curie or N\'eel temperature, therefore, provides a demanding test of theoretical descriptions of magnetic materials. A common computational approach maps \emph{ab initio} electronic-structure calculations onto a Heisenberg spin Hamiltonian and uses Monte Carlo simulations to determine its thermal behavior~\cite{archer2011exchange,mosleh2023_pbeu,rezaei2026benchmarking}. Discrepancies with experiments can arise from errors in the exchange interactions, limitations of the effective spin Hamiltonian, or the treatment of the spins as classical rather than quantum objects. Distinguishing these contributions is essential for understanding and improving predictive accuracy, because a computed ordering temperature reflects all of them at once.

The exchange interactions are most often obtained from density functional theory (DFT), but its results depend on the exchange-correlation approximation. Benchmarks of antiferromagnetic insulators have shown that the generalized gradient approximation (GGA) tends to overestimate ordering temperatures, whereas GGA+$U$ tends to underestimate them when the resulting spin models are treated classically~\cite{mosleh2023_pbeu}. SCAN~\cite{scan2015} and r$^2$SCAN~\cite{r2scan2020} meta-GGA functionals improve the overall agreement with experiment in a larger benchmark \cite{nafise_r2scan}. These comparisons are made at the level of transition temperatures; however, they do not separate the error in the exchange interactions from that in the treatment of the spins.

\begin{figure}[t]
    \centering
    \includegraphics[width=\linewidth]{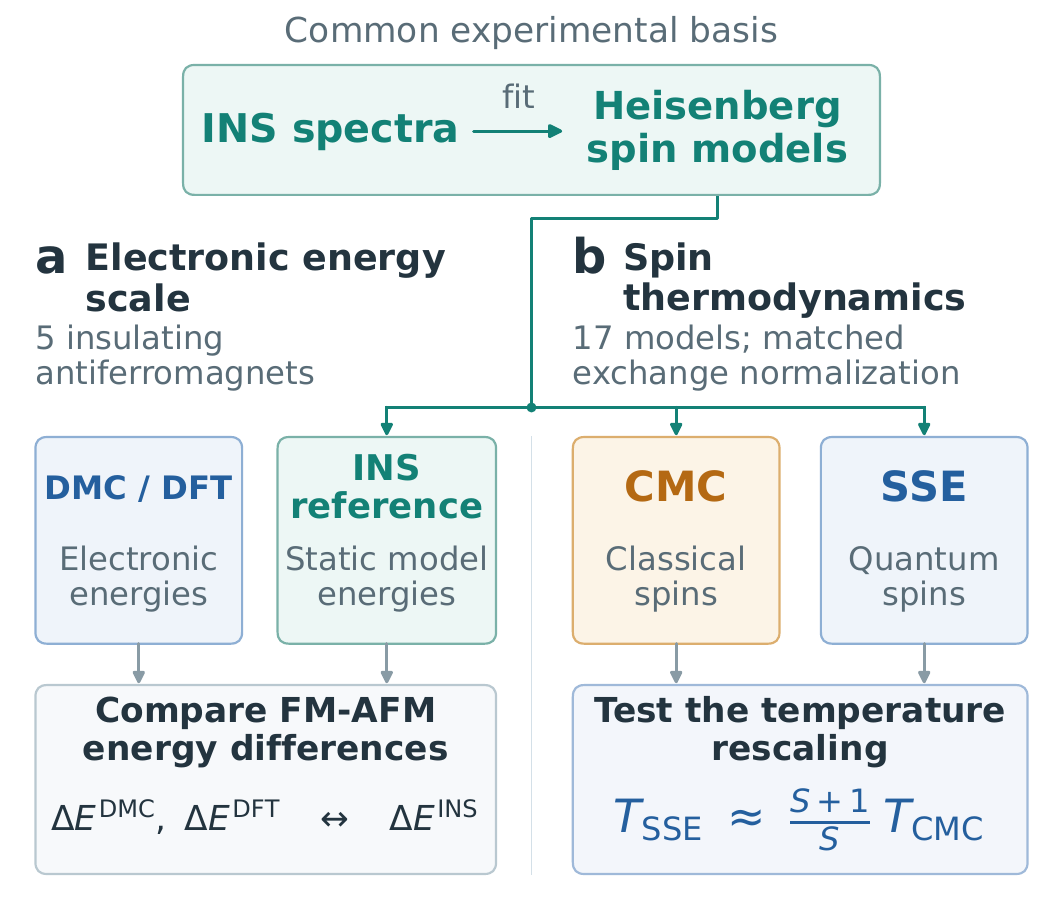}
\caption{\textbf{Complementary benchmarks of magnetic energy scales
and spin thermodynamics.} Exchange parameters fitted to INS spectra provide a common experimental basis.
\textbf{a}, FM--AFM energy differences
per magnetic atom obtained using electronic DMC and DFT are compared with static
energy differences reconstructed from INS-derived Heisenberg models
for five insulating antiferromagnets.
\textbf{b}, CMC and quantum SSE transition-temperature estimates are compared for
17 compounds: 14 AFM compounds, one FiM RbNiF$_3$,
and two FM compounds, CrI$_3$ and EuO. Both methods use the same
retained INS-derived interactions, with normalization conventions
that match collinear exchange energies. This comparison tests
$T_{\mathrm{SSE}}\approx[(S+1)/S]T_{\mathrm{CMC}}$, where $S$ is the
spin quantum number. The electronic and thermodynamic benchmarks
are performed separately; the SSE calculations use INS-derived
exchange parameters.}
    \label{fig:benchamrks}
\end{figure}

Figure~\ref{fig:benchamrks} summarizes our two complementary
benchmarks: magnetic energy differences that constrain the exchange
interactions (panel~(a)) and classical versus quantum transition
temperatures calculated from the corresponding spin Hamiltonians (panel~(b)). Errors in the electronic magnetic energy scale and
in the classical treatment of spins can compensate, so improving
the former may expose limitations of the latter.

The first of these two benchmarks concerns the electronic energy scale. Diffusion Monte Carlo (DMC) provides a complementary approach that treats electronic correlation explicitly. It has been applied to manganese oxides~\cite{mno_qmc_2018}, to the superexchange interaction in Ca$_2$CuO$_3$~\cite{ca2cuo3_dmc}, and to the magnetic properties of two-dimensional (2D) CrX$_3$ and VSe$_2$~\cite{crx3_qmc,vse2_dmc}. These studies motivate using DMC as an independent many-body reference for magnetic energy differences obtained from approximate density functionals, while recognizing its own trial-wave-function, pseudopotential, and convergence errors. A ferromagnetic--antiferromagnetic (FM--AFM) energy difference offers a targeted test of the magnetic energy scale. For systems with several exchange pathways, however, determining the individual exchange parameters generally requires additional magnetic configurations~\cite{mosleh2023_pbeu,rezaei2026benchmarking}, making a complete DMC-based parameterization substantially more demanding. We therefore restrict the electronic benchmark to the FM--AFM energy difference, using a reference reconstructed from exchange interactions derived from inelastic neutron scattering (INS) for the corresponding magnetic configurations.

The second benchmark concerns the spins themselves. Classical Monte Carlo (CMC) represents spins as continuous vectors, whereas a quantum spin model retains their finite-dimensional quantum states. CMC samples the Boltzmann distribution of a classical spin Hamiltonian and can treat frustrated exchange networks, single-ion anisotropy, and antisymmetric exchange, although equilibration can be demanding. For quantum spin models, stochastic series expansion (SSE) provides estimates with controlled statistical errors when efficient sampling is possible~\cite{syljuasen2002directed,alet2005directed}. Many frustrated Hamiltonians, however, produce a sign problem in the conventional spin basis, making simulations prohibitively expensive at low temperatures or large system sizes. The occurrence and severity of this problem depend on the Hamiltonian and its representation. In this study, we restrict the comparison to models that permit sign-problem-free SSE sampling, including reduced Hamiltonians for NiO and CrI$_3$ (Methods).

Where efficient quantum simulations are impractical, CMC remains a broadly applicable approach to spin thermodynamics. An approximation connecting classical and quantum transition temperatures would therefore extend its usefulness. Rescaling CMC transition temperatures by $(S+1)/S$, where $S$ is the spin quantum number, has been proposed for this purpose~\cite{nio_mno_alaei2023,mosleh2023_pbeu,alaei2025experimental}. This factor is motivated by the ratio of quantum and classical local-spin responses and the corresponding mean-field and isotropic random-phase-approximation predictions. Pavizhakumari and Olsen examined its applicability across several theoretical approaches~\cite{pavizhakumari2025neel}. Our previous studies assessed the $(S+1)/S$ rescaling primarily through comparisons with experimental ordering temperatures~\cite{nio_mno_alaei2023,alaei2025experimental}. Such comparisons also depend on how faithfully the fitted spin Hamiltonian represents the material. Here we test the rescaling directly against SSE simulations of corresponding quantum Hamiltonians, using the same retained INS-derived interactions in both classical and quantum calculations. This comparison evaluates the rescaling separately from discrepancies between the spin model and experiment.

Electronic DMC and lattice quantum Monte Carlo address successive stages in predicting magnetic ordering temperatures. DMC treats electronic many-body correlations when calculating magnetic configuration energies, which can be used to parameterize an effective spin Hamiltonian. SSE evaluates the finite-temperature quantum thermodynamics of that Hamiltonian. Here, we assess these stages separately: we benchmark electronic FM--AFM energy differences and test the previously proposed $(S+1)/S$ rescaling of CMC transition temperatures against SSE simulations using INS-derived spin Hamiltonians.

Exchange interactions fitted to INS spectra provide a common experimental basis for both benchmarks~\cite{nio_ins,mno_ins,alaei2025experimental,mnte_ins2006,cr2o3_ins1970,fe2o3_ins1970,wang2019local, Qu_2022}. 
They allow magnetic energy differences to be reconstructed for specified spin configurations and enable comparisons of classical and quantum thermodynamics without introducing errors from DFT-derived exchange parameters.

We use two partially overlapping material sets. The electronic benchmark covers five insulating antiferromagnets---NiO, MnO, Cr$_2$O$_3$, Fe$_2$O$_3$, and MnTe---for which we compare DMC and DFT FM--AFM energy differences with INS-based references. The thermodynamic benchmark uses INS-derived Hamiltonians for 17 compounds, including three of these five materials~\cite{alaei2025experimental}. This broader set is restricted to models that permit sign-problem-free SSE sampling. The reductions required for NiO and CrI$_3$ are described in Methods and applied consistently in CMC and SSE.

Here we find that DMC systematically underestimates the INS-based energy differences while achieving an overall error comparable to the best-performing density functionals tested. For the 17 spin models, the $(S+1)/S$ rescaling reduces the mean absolute difference between CMC and SSE transition-temperature estimates from 81.7 to 8.2~K. These complementary benchmarks assess electronic magnetic energy scales and support a practical approximation of quantum ordering temperatures within the tested spin models.


\section{Results}
\subsection{DMC wave-function optimization and convergence tests}
\begin{figure}[t]
    \centering
    \includegraphics[width=0.8\linewidth]{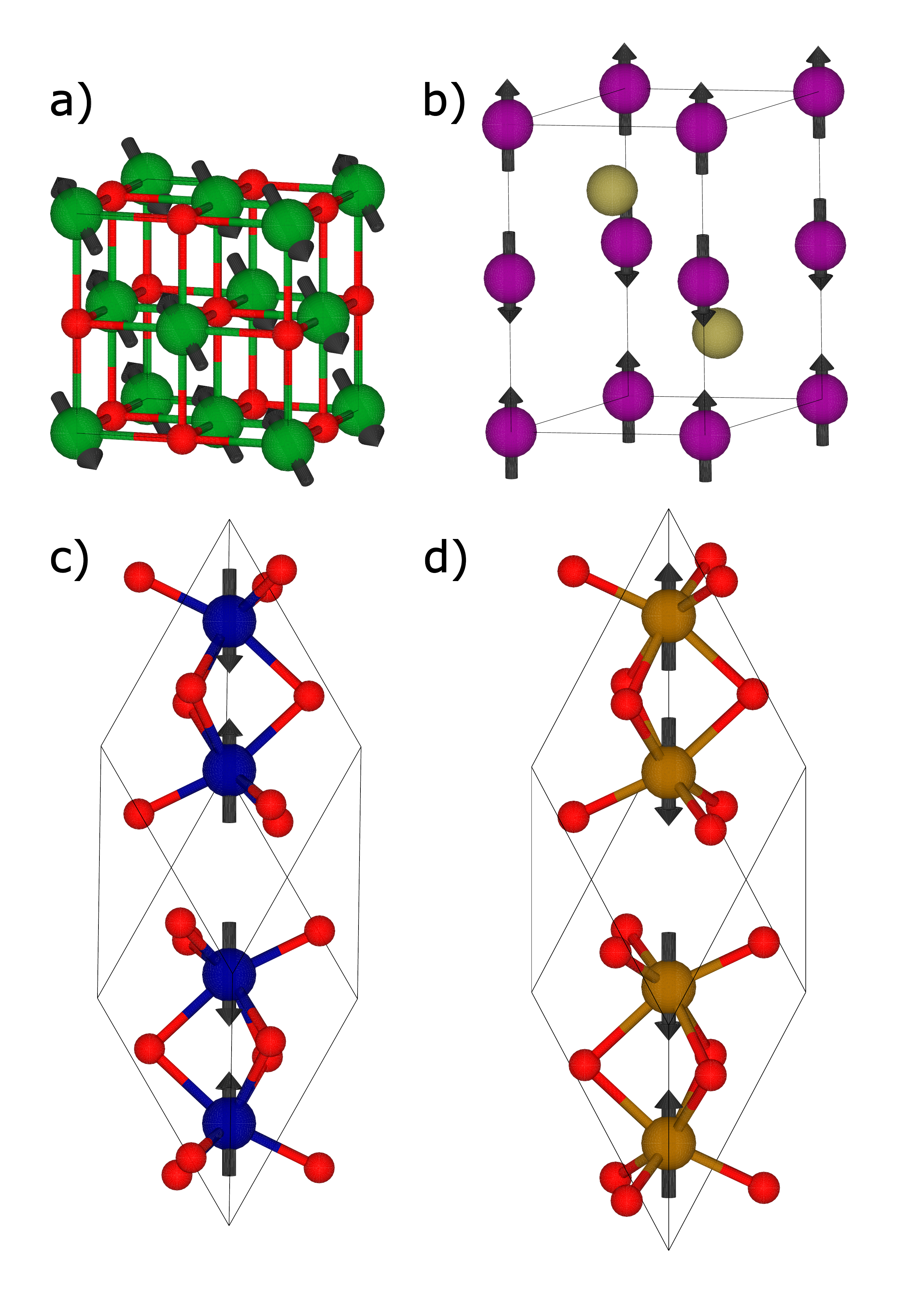}
    \caption{\textbf{
    Collinear AFM configurations of the benchmark compounds considered in DMC and DFT calculations
    } \textbf{a}, NiO; MnO has the same magnetic ordering pattern. \textbf{b}, MnTe. \textbf{c}, Cr$_2$O$_3$. \textbf{d}, Fe$_2$O$_3$. Green, purple, blue, and orange spheres represent Ni, Mn, Cr, and Fe atoms, respectively; red and yellow spheres represent O and Te atoms. Arrows indicate the relative orientations of the magnetic moments. The illustrated configurations represent the experimentally established AFM ordering patterns of the respective compounds, without accounting for magnetic anisotropy, which determines the orientation of the Néel vector with respect to the crystallographic axes.} 
    \label{fig::afm}
\end{figure}

Figure~\ref{fig::afm} shows the collinear AFM configurations used in the FM--AFM energy comparisons. The corresponding FM configurations were constructed by aligning all magnetic moments parallel.

For the five insulating compounds, we evaluate FM--AFM energy differences from fixed-node DMC total energies using PBE+$U$ trial orbitals. The convergence tests are detailed in the Methods subsection “Diffusion Monte Carlo calculations.”

\begin{figure*}
    \centering
    \includegraphics[width=1.0\linewidth]{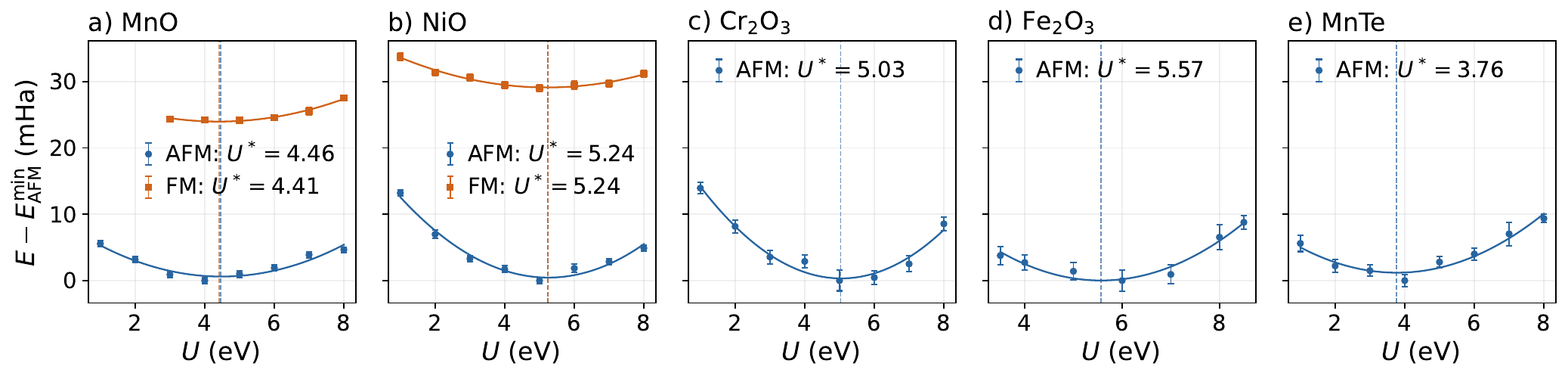}
    \caption{\textbf{Optimization of DMC trial orbitals.} DMC energies as a function of the Hubbard parameter $U$ used to generate the trial orbitals for \textbf{a}, MnO; \textbf{b}, NiO; \textbf{c}, Cr$_2$O$_3$; \textbf{d}, Fe$_2$O$_3$; and \textbf{e}, MnTe. Blue circles and orange squares denote AFM and FM configurations, respectively. For each material, energies per magnetic-structure cell are plotted relative to the lowest sampled AFM energy, with the same reference used for both magnetic configurations. Error bars indicate statistical uncertainties, solid curves show quadratic fits, and vertical dashed lines mark the fitted minima $U^*$. Metallic trial-orbital solutions at $U=1$ and 2~eV for FM MnO and at $U=1$, 2, and 3~eV for AFM Fe$_2$O$_3$ were excluded from the fits.}
    \label{uopt::mno_nio}
\end{figure*}

Figure~\ref{uopt::mno_nio} shows the DMC energy as a function of $U$, the Hubbard parameter in the PBE+$U$ calculations used to generate the DMC trial orbitals. Varying $U$ changes the orbitals and hence the nodal surface of the Slater determinant. The quadratic fits yield minima at $U^*_{\mathrm{AFM}}=4.46$~eV and $U^*_{\mathrm{FM}}=4.41$~eV for MnO, and at $U^*=5.24$~eV for both magnetic configurations of NiO. We define $U^*$ as the value of $U$ at the minimum of the fitted fixed-node DMC energy curve and use it to generate the production trial orbitals. The similar FM and AFM fitted minima suggest that $U^*$ depends only weakly on magnetic configuration in MnO and NiO. This trial-orbital optimization parameter is distinct from the linear-response parameter $U^{\mathrm{LR}}$ used in the DFT benchmark. The procedures for determining $U^{\mathrm{LR}}$ and $U^*$ are described in the Methods subsections “DFT calculations” and “Diffusion Monte Carlo calculations,” respectively.

For the AFM configurations of Cr$_2$O$_3$, Fe$_2$O$_3$, and MnTe, the fitted minima occur at approximately 5.03, 5.57, and 3.76~eV, respectively. Motivated by the similar FM and AFM minima in MnO and NiO, we used the AFM-optimized value for both magnetic configurations of these three compounds. This transfer remains an approximation because their FM minima were not independently determined.

\begin{figure}
    \centering
    \includegraphics[width=1.0\linewidth]{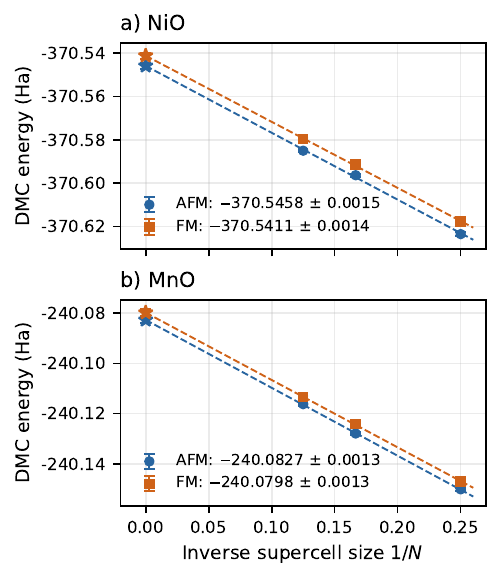}
    \caption{\textbf{Finite-size dependence of DMC energies.} Zero-time-step AFM and FM energies for \textbf{a}, NiO and \textbf{b}, MnO, plotted against inverse supercell size $1/N$. Here, $N=4$, 6, and 8 denotes the supercell multiplicity relative to the cell used for energy normalization. Energies are reported per magnetic-structure cell. Blue circles and orange squares denote AFM and FM configurations, respectively, and error bars indicate statistical uncertainties. Dashed lines show weighted linear fits, and stars mark the infinite-size extrapolations at $1/N=0$. The legends give the extrapolated energies and their statistical uncertainties in Ha.}
    \label{sizeffect}
\end{figure}


For NiO and MnO, the FM and AFM total energies show similar dependences on the DMC time step and supercell size (Fig.~\ref{sizeffect} and Supplementary Fig.~1S). The FM--AFM energy differences calculated with a time step of $0.005~\mathrm{Ha}^{-1}$ and supercells containing 16 magnetic atoms agree with the combined zero-time-step and infinite-size extrapolated estimates within statistical uncertainties (Supplementary Fig.~2S).
The larger uncertainties of the extrapolated differences limit the
precision with which residual time-step and finite-size biases can
be assessed. Equivalent convergence tests were not performed for
Cr$_2$O$_3$, Fe$_2$O$_3$, or MnTe.

For all five compounds, the FM--AFM energy differences remain consistent within statistical uncertainties when the number of discarded equilibration blocks is increased from 50 to 100 (Supplementary Fig.~3S).

\subsection{Benchmarking magnetic energy differences} 

Magnetic energy differences are reported per magnetic atom. For each material, the energy difference is calculated as
\begin{equation}
\Delta E^{\mathrm{M}}
=
\frac{
E_{\mathrm{FM,tot}}^{\mathrm{M}}
-
E_{\mathrm{AFM,tot}}^{\mathrm{M}}
}{N_{\mathrm{mag}}},
\label{eq:deltaE}
\end{equation}
where $\mathrm{M}={\mathrm{DFT},\mathrm{DMC}}$, $E_{\mathrm{FM,tot}}^{\mathrm{M}}$ and $E_{\mathrm{AFM,tot}}^{\mathrm{M}}$ denote the total energies of the FM and collinear AFM configurations, respectively, and $N_{\mathrm{mag}}$ is the number of magnetic atoms in the common simulation cell. The same FM and collinear AFM configurations are used to evaluate the DFT- and DMC-based energy differences, as well as the corresponding INS-based values.

\begin{figure}
    \centering
    \includegraphics[width=1.0\linewidth]{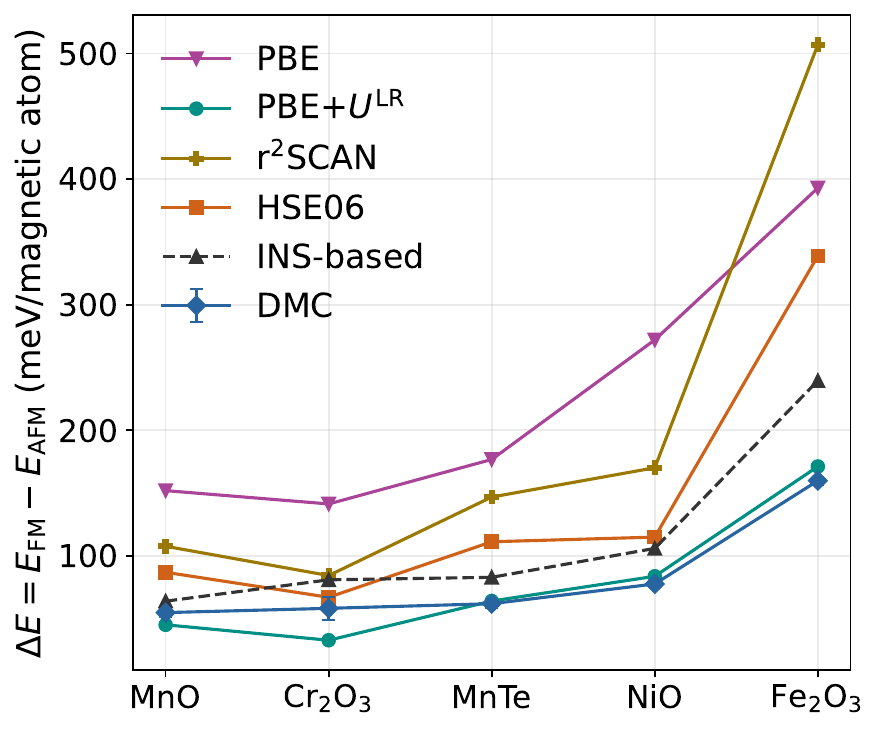}
    \caption{\textbf{Comparison of magnetic energy differences.} FM--AFM energy differences per magnetic atom obtained using PBE, PBE+$U^{\mathrm{LR}}$, r$^2$SCAN, HSE06, and DMC, together with the INS-based model references. Error bars indicate the statistical uncertainties of the DMC results. Lines connecting the compounds serve as guides to the eye.}
    \label{dqmc:comparison}
\end{figure}

The INS-based references are static energy differences reconstructed from INS-fitted Heisenberg models, as described in the Methods subsection “Calculation of INS-based reference energy differences.” These are model energies for the specified spin configurations, not directly measured electronic total-energy differences.
For PBE+$U^{\mathrm{LR}}$, $U^{\mathrm{LR}}$ is determined separately for the FM and AFM configurations using linear response through density-functional perturbation theory (DFPT); thus, the reported energy difference compares total energies calculated with their respective $U^{\mathrm{LR}}$ values. Details are given in the Methods subsection “DFT calculations.”
Table~\ref{diffs_tab} and Fig.~\ref{dqmc:comparison} compare the FM--AFM energy differences obtained using DMC and DFT with the INS-based reference values~\cite{alaei2025experimental}.  Every method yields a positive energy difference, placing the AFM configuration below the corresponding FM configuration.


\begin{table*}[t]
    \centering
    \caption{\textbf{Benchmark of FM--AFM energy differences.}
    Energy differences calculated using DFT and DMC and compared
    with INS-based model references. All values are in meV per
    magnetic atom. DMC entries include statistical uncertainties.
    Compounds are ordered by increasing INS-based reference energy
    difference. The mean absolute error (MAE) is evaluated over
    the five AFM compounds relative to the INS-based reference values.}
    \label{diffs_tab}

    \begin{tabular}{lrrrrrr}
        \toprule
        Compound & PBE & PBE+$U^{\mathrm{LR}}$
        & r$^2$SCAN & HSE06 & DMC & INS-based \\
        \midrule
        MnO
        & 152.0 & 45.2 & 107.7 & 87
        & $54.9 \pm 3.5$ & 64 \\

        Cr$_2$O$_3$
        & 141.4 & 32.9 & 84.4 & 67.1
        & $58.3 \pm 9.2$ & 81 \\

        MnTe
        & 176.8 & 64.1 & 146.9 & 111.3
        & $62.0 \pm 3.2$ & 83 \\

        NiO
        & 272.0 & 83.8 & 170.2 & 115
        & $77.7 \pm 2.6$ & 106 \\

        Fe$_2$O$_3$
        & 392.9 & 171.2 & 506.9 & 338.8
        & $159.9 \pm 3.3$ & 239.5 \\
        \midrule
        MAE
        & 112.3 & 35.3 & 88.5 & 34.7 & 32.1 & -- \\
        \bottomrule
    \end{tabular}
\end{table*}

DMC yields the smallest MAE in this benchmark, 32.1~meV per magnetic atom, followed closely by HSE06 and PBE+$U^{\mathrm{LR}}$ at 34.7 and 35.3~meV per magnetic atom, respectively. These three approaches therefore provide comparable average agreement with the reference values. PBE and r$^2$SCAN have larger MAEs of 112.3 and 88.5~meV per magnetic atom, respectively. The relative accuracy nevertheless depends on the compound: no method is closest to the reference in every case. For example, r$^2$SCAN gives the closest agreement for Cr$_2$O$_3$, despite its larger overall MAE.

The deviations also exhibit distinct systematic trends. PBE and r$^2$SCAN overestimate the reference energy differences for all five compounds, whereas DMC and PBE+$U^{\mathrm{LR}}$ underestimate them throughout the set. HSE06 overestimates the reference values for four compounds and underestimates the value for Cr$_2$O$_3$. The largest DMC deviation occurs for Fe$_2$O$_3$, where the calculated difference of $159.9\pm3.3$~meV per magnetic atom falls below the reference value of 239.5~meV per magnetic atom by approximately 33\%.

 For Fe$_2$O$_3$ relative to Cr$_2$O$_3$, the INS-based energy-difference ratio is 2.96. DMC and PBE yield ratios of 2.74 and 2.78, respectively, whereas HSE06, PBE+$U^{\mathrm{LR}}$, and r$^2$SCAN give 5.05, 5.20, and 6.01. For this pair, DMC reproduces the relative magnetic energy scale more closely than HSE06 and PBE+$U^{\mathrm{LR}}$, despite their similar overall MAEs. PBE also reproduces the ratio closely, although it substantially overestimates both individual energy differences.

Across the five compounds, larger INS-based FM--AFM energy differences are associated with higher experimental ordering temperatures (Supplementary Fig.~S6). This trend illustrates the relevance of the magnetic energy scale, although a single energy difference does not determine the transition temperature, which also depends on the exchange network and spin thermodynamics.

\subsection{Benchmarking classical and quantum transition temperatures}
The present temperatures are obtained from the approximately 4000-site campaign; parenthesized values retain the previously published repository results. Summary comparison statistics are evaluated from the unrounded sampled maxima, with the rescaling applied before rounding.

CMC and SSE use the same retained INS-derived Heisenberg interactions for each compound, with the exchange normalization chosen to match corresponding collinear-state energies. We report the finite-size, temperature-grid maxima of the specific heat (see Supplementary  Figs. 4S and 5S); full simulation details are given in the Methods subsection “Monte Carlo simulations of Heisenberg spin models.” NiO and CrI$_3$ use reduced Hamiltonians to permit sign-problem-free SSE sampling, and identical reductions were applied in CMC to ensure a consistent comparison. The omitted interactions are specified in the table note; CMC itself can treat the full interaction sets.

Table~\ref{tab:curie_temperatures} compares CMC and SSE transition-temperature estimates for 17 compounds using INS-derived exchange parameters. The set comprises 14 AFM compounds, one ferrimagnet
(FiM), RbNiF$_3$, and two FM compounds, CrI$_3$ and EuO.

Our previous studies found that multiplying CMC transition temperatures by $(S+1)/S$ improved agreement with experiment~\cite{nio_mno_alaei2023,alaei2025experimental}. A thermodynamic motivation for this factor follows from the weak-field response of an isotropic noninteracting spin: the quantum response is proportional to $S(S+1)$, whereas the response of a classical vector of length $S$ is proportional to $S^2$~\cite{andrade2020susceptibility}. Correspondingly, mean-field theory and the isotropic RPA expression give this ratio between their quantum and classical transition-temperature estimates, as discussed by Pavizhakumari and Olsen~\cite{pavizhakumari2025neel}. This motivates the rescaled CMC estimate
\begin{equation}
    T_{\mathrm{CMC}}^*
    =
    \frac{S+1}{S}T_{\mathrm{CMC}}.
    \label{eq:cmc_temperature_rescaling}
\end{equation}

\begin{figure*}[t]
    \centering
    \includegraphics[width=0.95\textwidth]{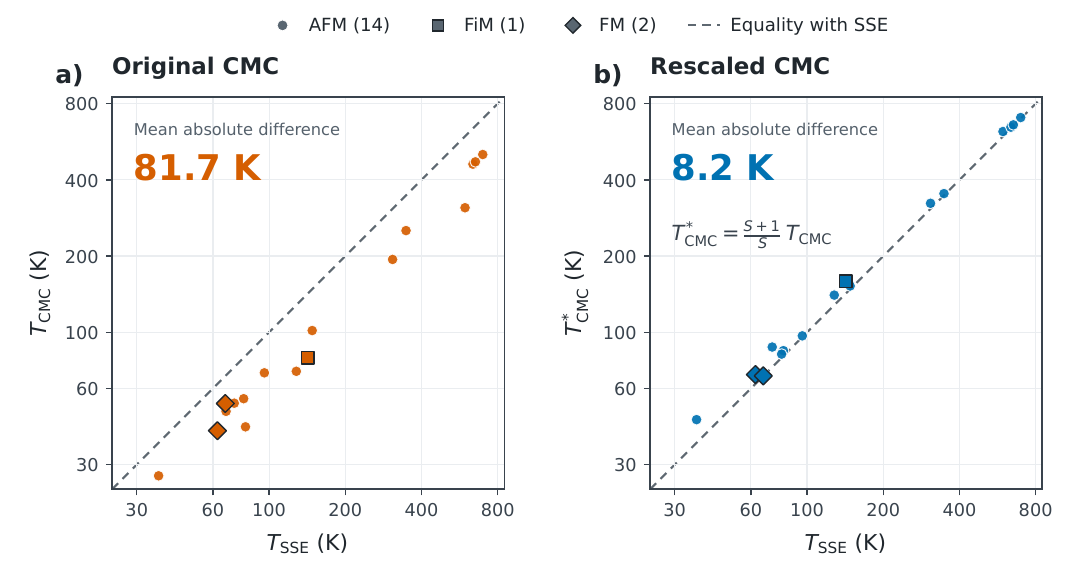}
    \caption{\textbf{Comparison of classical and quantum
    transition-temperature estimates.}
    \textbf{a}, Original CMC temperatures plotted against SSE estimates.
    \textbf{b}, Rescaled CMC temperatures,
    $T_{\mathrm{CMC}}^*=[(S+1)/S]T_{\mathrm{CMC}}$,
    plotted against the same SSE estimates, where $S$ is the spin
    quantum number. Circles represent the 14 AFM compounds, the square represents one FiM compound RbNiF$_3$, and
    diamonds represent the two FM compounds, CrI$_3$ and EuO.
    For each compound, CMC and SSE calculations use the same retained
    INS-derived exchange interactions. Dashed lines indicate equality
    with SSE; both panels use identical logarithmic axes.
    The mean absolute difference from SSE decreases from 81.7 to
    8.2~K after rescaling. The plotted values are the present results
    in Table~\ref{tab:curie_temperatures}, estimated from finite-size
    specific-heat maxima. Rescaling was applied before rounding.}
    \label{fig:cmc_sse_comparison}
\end{figure*}

The original CMC transition temperatures lie systematically below the
SSE estimates [Fig.~\ref{fig:cmc_sse_comparison}a]. Rescaling by
$(S+1)/S$ brings the temperatures close to the equality line
[Fig.~\ref{fig:cmc_sse_comparison}b], reducing the mean absolute
difference from 81.7 to 8.2~K across the 17 compounds. The rescaled
estimates have a mean absolute percentage difference of 5.9\%
relative to SSE. These statistics use the present CMC results in
Table~\ref{tab:curie_temperatures}; the parenthesized entries are
earlier results from the published exchange-interaction
dataset~\cite{alaei2025experimental}.

This agreement supports the rescaling beyond the mean-field and isotropic RPA arguments that motivate it. Despite its simplicity, the $(S+1)/S$ rescaling closely reproduces the SSE transition-temperature estimates across the tested models, with the remaining deviations discussed below. Because the exchange normalization already matches corresponding collinear-state energies, the rescaling concerns spin thermodynamics.

\begin{table}[]
    \centering
    \caption{\textbf{Classical and quantum magnetic transition
    temperatures.}
    CMC and SSE temperatures are estimated from finite-size
    specific-heat maxima. The first 14 compounds are AFM and
    are ordered by increasing spin quantum number $S$. The one FiM and two FM compounds
    are listed last.
    Experimental values denote N\'eel temperatures for the AFM
    compounds and Curie temperatures for the FiM and FM compounds.
    All temperatures are in kelvin and rounded to the nearest integer.
    The rescaled estimate is
    $T_{\mathrm{CMC}}^*=[(S+1)/S]T_{\mathrm{CMC}}$,
    with rescaling applied before rounding.
    Parenthesized entries are previously reported CMC temperatures
    and their rescaled values~\cite{alaei2025experimental}.}
    \label{tab:transition-temperatures}
    \label{tab:curie_temperatures}

    \normalsize
    \setlength{\tabcolsep}{2pt}
    \renewcommand{\arraystretch}{1.05}

    \begin{tabular}{@{}lccccc@{}}
        \toprule
        Compound & $S$ & $T_{\mathrm{CMC}}$
        & $T_{\mathrm{CMC}}^*$ & $T_{\mathrm{SSE}}$
        & $T_{\mathrm{exp}}$ \\
        \midrule

        LaVO$_3$     & 1
        & 70 (70) & 140 (140) & 128 & 150 \\
        NiF$_2$      & 1
        & 42 (42) & 85 (84) & 81 & 73 \\
        NiO          & 1
        & 311 (310) & 621 (620) & 595 & 523 \\
       
        \addlinespace[3pt]

        Cr$_2$O$_3$  & $3/2$
        & 194 (194) & 323 (323) & 308 & 308 \\
        Cr$_2$TeO$_6$ & $3/2$
        & 52 (53) & 87 (88) & 73 & 93 \\
        Cr$_2$WO$_6$ & $3/2$
        & 27 (27) & 45 (45) & 37 & 45 \\

        \addlinespace[3pt]

        FeF$_2$      & 2
        & 55 (55) & 82 (83) & 79 & 78 \\
        LaMnO$_3$    & 2
        & 102 (101) & 152 (152) & 148 & 140 \\

        \addlinespace[3pt]

        LaFeO$_3$    & $5/2$
        & 504 (504) & 705 (706) & 698 & 738 \\
        MnF$_2$      & $5/2$
        & 49 (48) & 68 (67) & 68 & 67 \\
        MnTe         & $5/2$
        & 252 (251) & 353 (351) & 348 & 310 \\
        NdFeO$_3$    & $5/2$
        & 461 (463) & 646 (648) & 639 & 693 \\
        PrFeO$_3$    & $5/2$
        & 471 (471) & 660 (659) & 653 & 707 \\
        RbMnCl$_3$   & $5/2$
        & 69 (69) & 97 (97) & 96 & 94 \\

   \midrule

        RbNiF$_3$ (FiM)    & 1
        & 79 (79) & 159 (158) & 142 & 133 \\
        \midrule

        CrI$_3$ (FM) & $3/2$
        & 41 (39) & 68 (65) & 63 & 61 \\
        EuO (FM)     & $7/2$
        & 52 (52) & 67 (67) & 67 & 69 \\

        \bottomrule
    \end{tabular}

    \vspace{3pt}
    \begin{minipage}{\linewidth}
        \footnotesize
        The present NiO model retains only $J_2$ and omits
        single-ion anisotropy; the present CrI$_3$ model omits $J_6$.
        Parenthesized values and experimental temperatures are taken
        from the published exchange-interaction
        dataset~\cite{alaei2025experimental}.
    \end{minipage}
\end{table}

At full precision, the rescaled estimates exceed the SSE values for all 17 compounds. At the integer precision reported in the table, 15 exceed the SSE values and two coincide, MnF$_2$ and EuO. The largest relative deviations occur for Cr$_2$WO$_6$ and Cr$_2$TeO$_6$, at approximately 23\% and 20\%, respectively, evaluated before rounding. For the two FM compounds, the rescaled CMC and SSE estimates are 68 and 63~K for CrI$_3$, and 67 and 67~K for EuO. The rescaling therefore improves agreement across both AFM and FM examples, while retaining material-dependent deviations.

Agreement with SSE must also be distinguished from agreement with experiment. The rescaled CMC and SSE estimates have similar mean absolute errors relative to experiment, 21.5 and 19.9~K, respectively. For MnTe, for example, the two estimates are close, 353 and 348~K, but both exceed the experimental value of 310~K. Reproducing the quantum result of the chosen spin model therefore does not eliminate discrepancies between that model and the material. Overall, the comparison supports $(S+1)/S$ as a practical approximation for the selected Hamiltonians, subject to the finite-size and temperature-grid limitations described in Methods.

\section{Discussion}
Our results emphasize that reliable magnetic transition temperatures require both an accurate electronic magnetic energy scale and an appropriate treatment of quantum spin thermodynamics. Improving either stage should be assessed independently, because agreement between a calculated ordering temperature and experiment can conceal an inaccurate description of the underlying magnetic interactions.

The magnetic energy differences provide a useful criterion for this assessment. PBE substantially overestimates the INS-based FM--AFM energy differences, whereas DMC yields smaller values than the references. HSE06 and PBE+$U^{\mathrm{LR}}$ provide average agreement comparable to DMC, although their deviations differ among compounds. These differences demonstrate the sensitivity of magnetic energetics to the electronic-structure approximation. Benchmarking competing magnetic configurations therefore offers a targeted way to evaluate exchange-correlation functionals and guide their improvement. Extending the comparison beyond a single FM--AFM difference would constrain additional combinations of exchange interactions and provide a more complete assessment of the exchange network.

The INS-based reference energies are reconstructed from effective Heisenberg models fitted to magnetic excitation spectra. Using these models to evaluate static FM--AFM energy differences assumes that the fitted Hamiltonians also describe the energies of these configurations. A further consideration is the correspondence between electronic magnetic states and spin-model states. Torelli and Olsen demonstrated, for two-dimensional magnets, that accounting for quantum correlations in this mapping can modify the extracted exchange parameters~\cite{torelli2020quantum}. Discrepancies with electronic-structure calculations can therefore reflect errors in the calculated energies, uncertainties in the fitted exchanges, or limitations of the effective Hamiltonian and the mapping procedure.

The connection between magnetic energy scales and spin thermodynamics gives these benchmarks practical significance. An excessive exchange-energy scale can raise transition temperatures obtained from CMC, as expected from PBE's overestimated magnetic energy differences. DMC instead underestimates the INS-based energy differences. If this smaller energy scale carries over to the extracted exchanges, a classical treatment would introduce a further underestimate relative to the corresponding quantum spin model. Thus, obtaining more reliable magnetic energies from DMC or DFT
should be accompanied by explicit consideration of quantum spin thermodynamics. The suitability of unrescaled CMC cannot be inferred from the sophistication of the electronic method alone.

Our direct CMC--SSE comparison demonstrates a practical way to address this issue. For the corresponding INS-derived Hamiltonians, multiplying CMC temperatures by $(S+1)/S$ closely reproduces the SSE estimates across the AFM, FiM, and FM models studied here. Because the classical and quantum models share a common exchange-energy normalization, this agreement tests the temperature rescaling separately from errors in electronically calculated exchanges. It supports combining accurate exchange parameters with quantum spin simulations where feasible, or with the rescaling as an approximation supported by the present benchmark. Agreement with experiment additionally depends on the adequacy of the chosen spin Hamiltonian.

DMC and complementary quantum-chemical methods can help develop the electronic benchmarks needed for this strategy. Booth \emph{et al.} demonstrated the use of FCIQMC reference energies to assess quantum-chemical methods in solids, including calculations of the NiO spin gap~\cite{booth2013exactsolids}. Such comparisons show the value of applying independent many-electron approaches to common magnetic systems. Broader comparisons of magnetic energy differences obtained using DMC, quantum-chemical methods, and density functionals could help identify reliable descriptions of exchange interactions and establish reference datasets for functional development.

These findings motivate a prediction strategy that first benchmarks magnetic energy differences, then determines the relevant exchange interactions, and finally addresses quantum spin thermodynamics when calculating ordering temperatures. Here, the electronic and thermodynamic stages are tested separately. Extending this approach to additional magnetic configurations, smaller-spin systems, and more complex exchange networks would help establish how broadly the $(S+1)/S$ approximation can complement accurate electronic calculations, particularly where efficient quantum spin simulations are difficult.

\section{Methods}\label{sec:methods}
\subsection{DFT calculations}\label{subsec:dft}
We performed spin-polarized DFT calculations for the FM and collinear AFM configurations of NiO, MnO, Cr$_2$O$_3$, Fe$_2$O$_3$, and MnTe using experimental lattice parameters and internal atomic coordinates. The lattice parameters are listed in Supplementary Table~1S.

The PBE~\cite{pbe1996},
HSE06~\cite{hse06paper,hse06_krukau2006}, and
r$^2$SCAN~\cite{r2scan2020} calculations were performed
using the Vienna Ab initio Simulation Package
(VASP)~\cite{vasp_paper} with the projector augmented-wave
(PAW) method~\cite{paw_paper}. The plane-wave kinetic-energy cutoffs were 550~eV for PBE, 500~eV for HSE06, and 600~eV for r$^2$SCAN. For PBE, the electronic convergence threshold was $10^{-6}$~eV. The PBE, HSE06, and r$^2$SCAN calculations used Monkhorst--Pack $k$-point meshes of $8\times8\times8$ for NiO and MnO, $6\times6\times6$ for Cr$_2$O$_3$ and Fe$_2$O$_3$, and $6\times6\times4$ for MnTe.

The PBE+$U^{\mathrm{LR}}$ calculations were performed using Quantum ESPRESSO~\cite{qe2017} with Garrity--Bennett--Rabe--Vanderbilt ultrasoft pseudopotentials~\cite{gbrv2014}. The Hubbard parameters $U^{\mathrm{LR}}$ were determined
self-consistently using linear response~\cite{cococcioni2005linear}
within DFPT, as implemented
in the \texttt{hp.x} code~\cite{timrov2018dfpt,timrov2022hp}. For each material, $U^{\mathrm{LR}}$ was calculated separately for the FM and AFM configurations and used in the corresponding total-energy calculation. For each magnetic configuration, the Hubbard parameter was updated iteratively: a DFPT linear-response calculation was performed using the electronic state obtained with the preceding $U$, and the resulting value was used in the next DFT+$U$ calculation. Iterations continued until successive values of $U^{\mathrm{LR}}$ differed by less than $0.0001$~eV. Thus, the PBE+$U^{\mathrm{LR}}$ energy differences compare total energies evaluated with the respective self-consistent Hubbard parameters. The kinetic-energy cutoffs for the wave functions and charge density were 50 and 500~Ry, respectively.

The PBE+$U^{\mathrm{LR}}$ benchmark calculations are separate from the Quantum ESPRESSO calculations used to generate the DMC trial orbitals. In particular, $U^{\mathrm{LR}}$ is determined from linear response, whereas the parameter $U^*$ used to generate the DMC trial orbitals is selected by minimizing the DMC energy, as described in the following subsection.

\subsection{Diffusion Monte Carlo calculations}\label{subsec:dmc}
We performed fixed-node DMC calculations using QMCPACK 4.1.0~\cite{kim2018qmcpack,kent2020qmcpack}, with workflows managed by Nexus~\cite{krogel2016nexus}. For each material, the FM and collinear AFM total energies were used to evaluate $\Delta E^{\mathrm{DMC}}$ according to Eq. \eqref{eq:deltaE}.

The trial wave function had the Slater--Jastrow form
\begin{equation}
    \Psi_{\mathrm{T}}(\mathbf{R})
    =
    e^{\Phi(\mathbf{R})}
    D_{\uparrow}(\mathbf{R})
    D_{\downarrow}(\mathbf{R}),
    \label{eq:trial_wavefunction}
\end{equation}
where $\mathbf{R}$ denotes the electronic coordinates, $D_{\uparrow}$ and $D_{\downarrow}$ are Slater determinants of spin-up and spin-down orbitals, and $\Phi$ contains electron--electron, electron--ion, and electron--electron--ion correlation terms. The two-body Jastrow functions were represented by B-splines and the three-body functions by polynomials. The Jastrow parameters were optimized in variational Monte Carlo using the OneShift optimization procedure, with 22 optimization iterations~\cite{kent2020qmcpack}.

The single-particle orbitals were generated using PBE+$U$~\cite{pbe1996,anisimov1997,dudarev98} in Quantum ESPRESSO 7.4.1~\cite{qe2017}, with the Hubbard correction applied to the transition-metal $3d$ states. Norm-conserving pseudopotentials were used for all elements~\cite{ni_mn_pp2022,o_pp2017,te_pp_2022}. The plane-wave kinetic-energy cutoff was 400~Ry for NiO, MnO, Cr$_2$O$_3$, and MnTe, and 350~Ry for Fe$_2$O$_3$. Brillouin-zone sampling employed Gamma-centered $k$-point meshes of $8\times8\times8$ for NiO, MnO, Cr$_2$O$_3$, and Fe$_2$O$_3$, and $10\times10\times6$ for MnTe. Nonlocal pseudopotential contributions in the DMC calculations were treated using the T-moves scheme~\cite{casula2006tmoves}.

The parameter $U$ used to generate the trial orbitals was varied to optimize the trial wave function within the chosen Slater--Jastrow family. For each magnetic configuration considered in this optimization, the value $U^*$ was determined from the minimum of a quadratic fit to the DMC energy as a function of $U$. For FM MnO and AFM Fe$_2$O$_3$, points whose orbital-generating calculations converged to metallic solutions were excluded from the fits. The FM and AFM configurations of NiO and MnO were optimized separately. For Cr$_2$O$_3$, Fe$_2$O$_3$, and MnTe, the AFM-optimized value was used for both configurations. The fitted values $U^*$ were then used to generate the orbitals for the production DMC calculations. 

Time-step and finite-size effects were assessed for NiO and MnO using three supercell sizes and time steps of 0.02, 0.01, and 0.005~Ha$^{-1}$. At each supercell size, the FM and AFM total energies were extrapolated linearly to zero time step. The resulting energies were subsequently extrapolated against inverse supercell size. These extrapolations provided reference estimates for assessing the FM--AFM energy differences obtained with the production settings.

The production calculations used supercells containing 16 magnetic atoms and a time step of 0.005~Ha$^{-1}$. Twist-averaged boundary conditions were used to reduce finite-size effects~\cite{lin2001twist}, with 36 unique twists for NiO and MnO, 34 for MnTe, and 24 for Cr$_2$O$_3$ and Fe$_2$O$_3$. The target walker populations were 2000 for MnO, Cr$_2$O$_3$, and MnTe, and 1500 for NiO and Fe$_2$O$_3$. Each DMC block comprised 40 propagation steps. The production runs used 200 blocks for MnO, Cr$_2$O$_3$, and MnTe, and 300 blocks for NiO and Fe$_2$O$_3$, corresponding to 8000 and 12000 propagation steps, respectively. The first 50 blocks---equivalent to 2000 propagation steps---were discarded as equilibration in all production runs. Sensitivity to the equilibration cutoff was checked by increasing the number of discarded blocks from 50 to 100. Statistical uncertainties in the FM--AFM energy differences were obtained by combining the uncertainties of the energies per magnetic atom in quadrature, assuming independent FM and AFM estimates.

\subsection{Monte Carlo simulations of Heisenberg spin models}\label{subsec:spinmc}

We compared CMC and SSE transition-temperature estimates for the 17 compounds listed in Table~\ref{tab:curie_temperatures}, using INS-derived exchange parameters compiled in Ref.~\cite{alaei2025experimental}. The quantum simulations employed the directed-loop SSE
implementation in
ALPS~\cite{bauer2011alps,syljuasen2002directed,alet2005directed},
\texttt{dirloop\_sse}, while the classical simulations used
the \texttt{spinmc} application with local updates. The spin quantum number $S$ assigned to each compound is given in the table.

The classical Hamiltonian was expressed in terms of unit vectors $\mathbf{e}_i$ as
\begin{equation}
    H_{\mathrm{cl}}
    =
    -\sum_{\langle i,j\rangle}
    J_{ij}^{\mathrm{cl}}\,\mathbf{e}_i\cdot\mathbf{e}_j
    +
    \sum_i D_i^{\mathrm{cl}}(e_i^z)^2,
    \label{eq:classical_spin_model}
\end{equation}
where $J_{ij}^{\mathrm{cl}}$ and $D_i^{\mathrm{cl}}$ are the exchange and single-ion anisotropy parameters, respectively. The corresponding quantum Hamiltonian was
\begin{equation}
    H_{\mathrm{q}}
    =
    \sum_{\langle i,j\rangle}
    J_{ij}^{\mathrm{q}}\,
    \hat{\mathbf{S}}_i\cdot\hat{\mathbf{S}}_j
    +
    \sum_i D_i^{\mathrm{q}}(\hat{S}_i^z)^2,
    \label{eq:quantum_spin_model}
\end{equation}
where $\hat{\mathbf{S}}_i$ is a spin-$S$ operator and $\hbar=1$. In both Hamiltonians, the exchange sum includes all retained bonds, with each bond counted once.

We used the parameter conversion
$J_{ij}^{\mathrm{q}}=-J_{ij}^{\mathrm{cl}}/S^2$ and
$D_i^{\mathrm{q}}=D_i^{\mathrm{cl}}/S^2$.
With these sign conventions, positive $J_{ij}^{\mathrm{cl}}$ favors parallel alignment, whereas positive $J_{ij}^{\mathrm{q}}$ favors antiparallel alignment. With this normalization, the exchange contributions to the energies of corresponding collinear classical configurations and quantum spin product states are equal. This establishes a common exchange-energy scale for comparing classical and quantum spin thermodynamics.

The material set was selected to permit SSE sampling without a sign problem. Reduced Hamiltonians were used for NiO and CrI$_3$. For NiO, only $J_2$ was retained; $J_1$ and the single-ion anisotropy were omitted. For CrI$_3$, $J_6$ was omitted, while $J_1$, $J_2$, and $J_3$ were retained. For each compound, the same retained interactions were used in the CMC and SSE calculations.

Both methods used the same periodic supercell for each compound, containing approximately 4000 magnetic sites (3840--4056 sites, depending on the compound). Each method sampled 60 temperatures. At every temperature, eight independently seeded simulations were performed, each comprising $10^6$ thermalization sweeps followed by at least $10^6$ measurement sweeps.

The specific heat was evaluated from energy fluctuations in CMC and from the expansion-order estimator in SSE. At each temperature, the reported specific heat is the mean of the eight independent-chain estimates. Its error bar is the larger of the replica standard error and the propagated within-chain paired-bin jackknife error, preserving the covariance of the energy or expansion-order moments. These error bars quantify uncertainty in the specific heat, not confidence intervals for the peak temperature. For each model and method, the transition temperature was estimated as the sampled temperature at which the specific heat was largest. These values are finite-size estimates obtained directly from the temperature grid; no peak fitting, interpolation, or extrapolation to the thermodynamic limit was performed.

The classical simulations used the uncorrected unit-vector Hamiltonian in equation~\eqref{eq:classical_spin_model}. After extracting the transition temperatures, we evaluated the rescaled estimates
$T_{\mathrm{CMC}}^*=[(S+1)/S]T_{\mathrm{CMC}}$
and compared them with $T_{\mathrm{SSE}}$. The rescaling was applied before rounding, and all reported temperatures were rounded to the nearest kelvin, with half ties rounded upward.

\subsection{Calculation of INS-based reference energy differences}\label{subsec:ins}
For the five compounds in the electronic-structure benchmark, we constructed reference FM--AFM energy differences using exchange parameters fitted to INS spectra and compiled in Ref.~\cite{alaei2025experimental}. The INS-based reference energy differences were evaluated using the full exchange parameterizations adopted for each compound. No interactions were removed for this calculation; the Hamiltonian reductions described above apply only to the CMC--SSE comparison. 
The original INS studies are those for NiO~\cite{nio_ins},
MnO~\cite{mno_ins}, Cr$_2$O$_3$~\cite{cr2o3_ins1970},
Fe$_2$O$_3$~\cite{fe2o3_ins1970}, and
MnTe~\cite{mnte_ins2006}.
The exchange parameters, crystal structures, and bond-resolved interaction data are available in the associated \texttt{Magnetic\_data} repository.

We used the exchange term of equation~\eqref{eq:classical_spin_model}, with the couplings expressed in the same unit-vector convention. For a collinear magnetic configuration $X$, the exchange energy per magnetic atom is
$E_X^{\mathrm{INS}}=-\sum_n\alpha_n^X J_n^{\mathrm{cl}}$,
where $J_n^{\mathrm{cl}}$ denotes the coupling for exchange shell $n$. The dimensionless coefficient
$\alpha_n^X=N_{\mathrm{mag}}^{-1}\sum_{\langle i,j\rangle_n}
\mathbf{e}_i^X\cdot\mathbf{e}_j^X$
accounts for the bond multiplicities and relative spin orientations, with each bond counted once. We evaluated these coefficients using the \textsc{ESpinS} code~\cite{espins2022} for the same FM and AFM configurations used in the DFT and DMC calculations.

The resulting reference energy difference per magnetic atom is
\begin{equation}
    \Delta E^{\mathrm{INS}}
    =
    E_{\mathrm{FM}}^{\mathrm{INS}}
    -
    E_{\mathrm{AFM}}^{\mathrm{INS}}
    =
    \sum_n
    \left(
        \alpha_n^{\mathrm{AFM}}
        -
        \alpha_n^{\mathrm{FM}}
    \right)
    J_n^{\mathrm{cl}},
    \label{dqmc::ins}
\end{equation}
where the sum runs over all exchange shells included in the adopted INS parameterization. The single-ion anisotropy term in equation~\eqref{eq:classical_spin_model} cancels between collinear FM and AFM configurations defined along the same axis. Evaluating these static energies requires no Monte Carlo sampling and is therefore not subject to the sign-problem restrictions of the SSE calculations.

These reference values are model energies evaluated for the specified collinear spin configurations, rather than directly measured FM--AFM electronic total-energy differences. No explicit quantum-ground-state correlation-energy correction is added.

\section*{Data availability}
The SSE quantum Monte Carlo and CMC spinmc simulation data supporting the transition-temperature analysis are available in the \texttt{Magnetic\_data} repository at \url{https://github.com/malaei/Magnetic_data/tree/main/INS_SW_MC/QMC}. The underlying DFT and DMC calculation data are not publicly available but may be obtained from the corresponding author on reasonable request.

\section*{Code availability}
The calculations used VASP, Quantum ESPRESSO, QMCPACK, and ALPS, as cited in Methods. Custom scripts used to prepare and run the DMC calculations and to analyze the data are available from the corresponding author upon reasonable request.

\clearpage
\providecommand{\bibcommenthead}{}
\bibliographystyle{sn-nature-linked}
\bibliography{literature}

\section*{Acknowledgements}
M.A. thanks Nafise Rezaei for helpful comments on the manuscript.

\subsection*{Use of AI-assisted tools}
During manuscript preparation, the authors used ChatGPT (OpenAI) to assist with language editing, manuscript organization, checks of scientific exposition and references, and preparation of Python code for figures. The authors reviewed and verified the resulting material and take full responsibility for the manuscript.

\section*{Funding}
A.Q. was supported by the Research Council of Norway through Grant Nos. 353919 and 361800 ``QTransMag'', and Grant No. 262633 ``QuSpin''.

\section*{Author contributions}
I.E.M. performed the DMC calculations. M.A. performed the CMC and QMC simulations and supervised the project. A.Q. contributed to the conceptual development of the study and the interpretation of the results. I.E.M. prepared the initial manuscript draft, which A.Q. and M.A. revised and edited. A.R.O. supported and facilitated the project. All authors participated in scientific discussions, contributed to the final revision of the manuscript, and approved its final version.

\section*{Competing interests}
The authors declare no competing interests.
\end{document}


\maketitle

\noindent
This Supplementary Information provides the structural parameters used in the calculations; additional DMC time-step, finite-size, and equilibration analyses; specific-heat curves from stochastic series expansion and classical Monte Carlo simulations for the 17 compounds studied; and a comparison of INS-based magnetic energy differences with experimental ordering temperatures. The specific-heat curves show the sampled finite-size maxima used to estimate the transition temperatures reported in the main text.
\par\vspace{1.5em}

\section*{Structural parameters}
\begin{table}[htbp]
    \centering
    \caption{Structure parameters of systems used in this study: chemical formula, lattice constants and space groups.}
    \begin{tabular}{ccc}
        \hline
        Material & Lattice parameters,~\AA & Space group \\
        \hline
        MnO~\cite{mno_qmc_2018} & $a=4.43$ & 225 ($Fm\overline{3}m$) \\
        NiO~\cite{nio_qmc_2017} & $a=4.17$ & 225 ($Fm\overline{3}m$) \\
        Cr$_2$O$_3$ & $a=4.96$, $c=13.56$ & 167 ($R\overline{3}c$) \\
        Fe$_2$O$_3$ & $a=5.04$, $c=13.76$ & 167 ($R\overline{3}c$) \\
        MnTe & $a=4.16$, $c=6.71$ & 194 ($P6_3/mmc$) \\
        \hline
    \end{tabular}
    \label{tab_struct}
\end{table}

\clearpage
\section*{DMC convergence tests}
\begin{figure}[htbp]
    \centering
    \includegraphics[width=\linewidth]{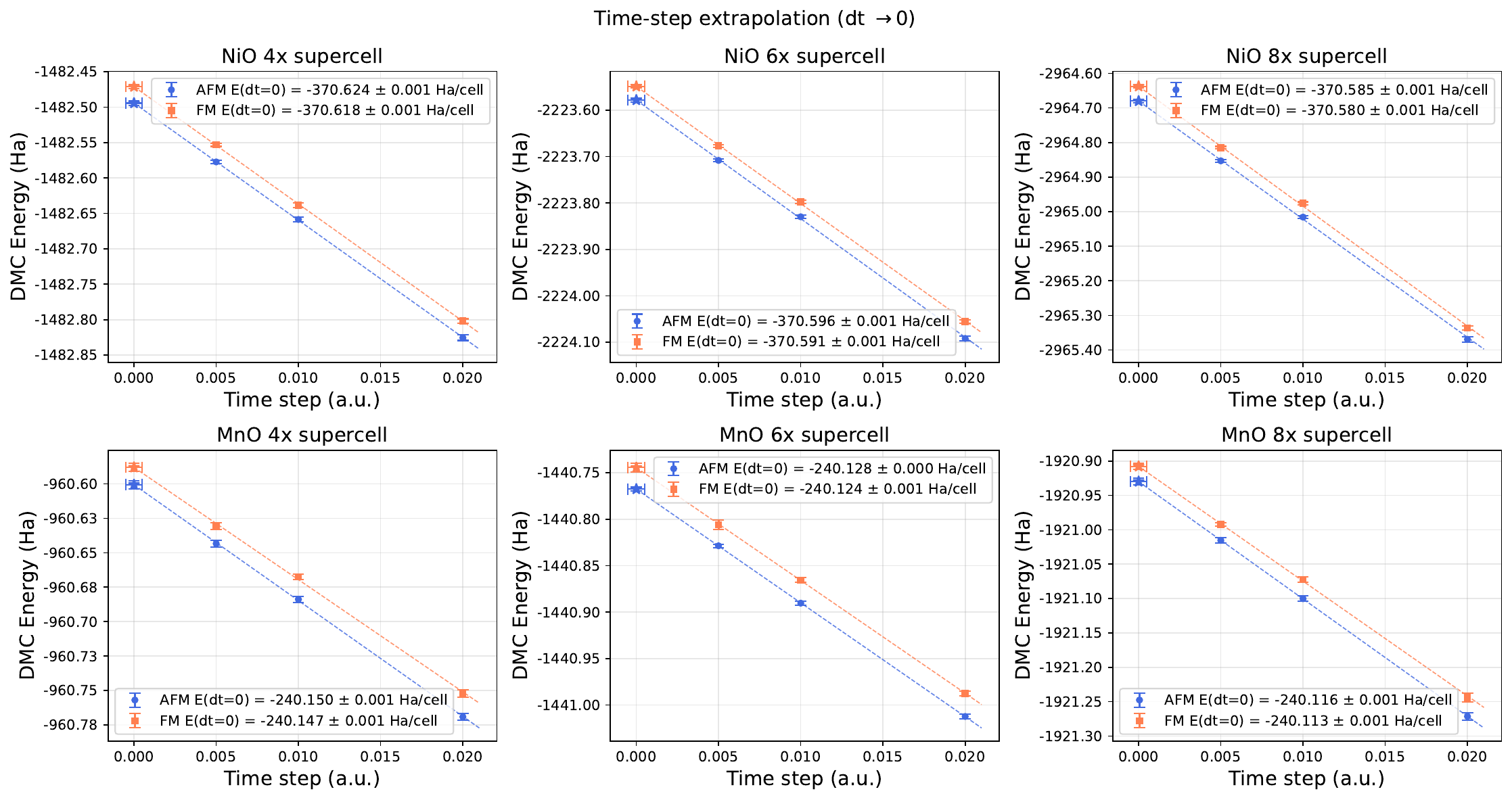}
    \caption{Time-step dependence and linear extrapolation of the FM and AFM DMC energies for three supercell sizes of NiO (top) and MnO (bottom). Points and fitted curves show total supercell energies; the zero-time-step values in the legends are normalized per magnetic-structure cell.}
    \label{fig:time_step_supplement}
\end{figure}

\begin{figure}[htbp]
    \centering
    \includegraphics[width=0.75\linewidth]{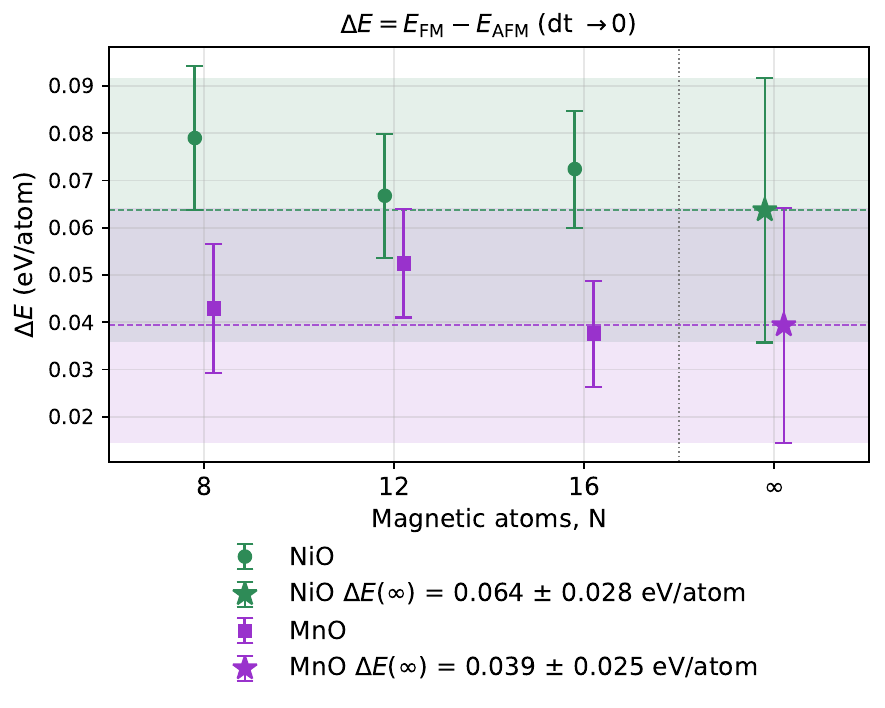}
    \caption{NiO and MnO DMC energy differences ($dt \to 0$) with their statistical uncertainties vs number of magnetic atoms in the supercell; $\infty$ indicates the extrapolated value.}
    \label{fig:de_vs_n_supplement}
\end{figure}

\begin{figure}[htbp]
    \centering
    \includegraphics[width=0.75\linewidth]{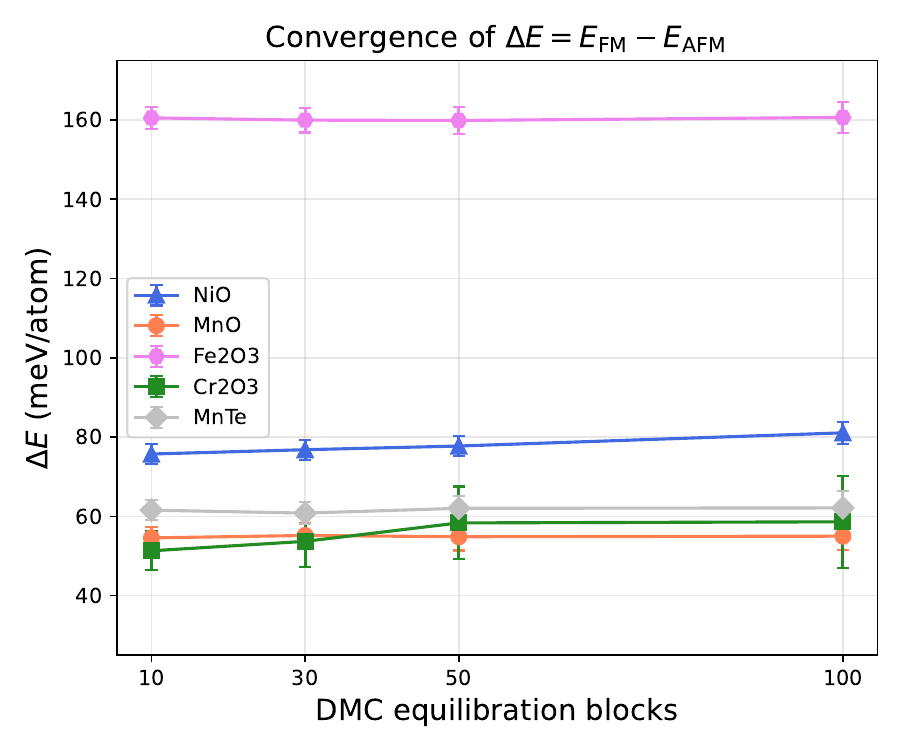}
    \caption{DMC energy differences as a function of the number of discarded equilibration blocks.}
    \label{fig:equilibration_supplement}
\end{figure}

\clearpage
\section*{Specific-heat curves}
The specific-heat curves and the transition-temperature estimates reported in the main text are obtained from the same approximately 4000-site simulations, with 3840--4056 magnetic sites depending on the compound. Vertical dotted lines indicate the largest sampled mean specific heat. All 34 CMC and SSE maxima lie inside the sampled temperature ranges. Curves show the mean of eight independent simulations at each temperature. Error bars are the larger of the replica standard error and the propagated within-chain paired-bin jackknife error; they quantify uncertainty in the specific heat, not confidence intervals for the peak temperature. The peak temperatures are selected directly from the grids, without interpolation or extrapolation to the thermodynamic limit.
\clearpage
\begin{landscape}
\begin{figure}[p]
    \centering
    \includegraphics[width=0.92\linewidth]{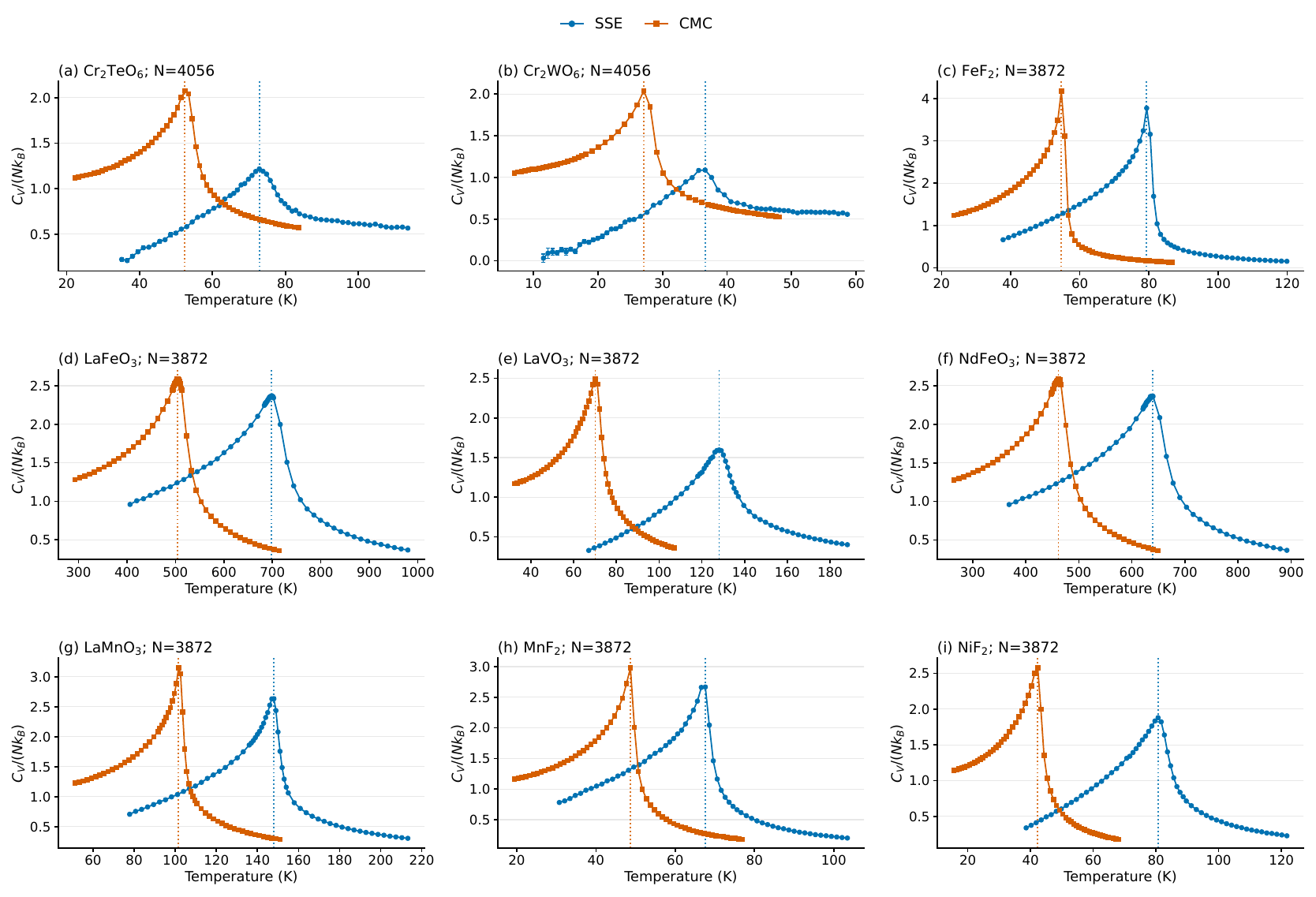}
    \caption{Specific heat per magnetic site from the approximately 4000-site SSE and CMC simulations for (a) Cr$_2$TeO$_6$, (b) Cr$_2$WO$_6$, (c) FeF$_2$, (d) LaFeO$_3$, (e) LaVO$_3$, (f) NdFeO$_3$, (g) LaMnO$_3$, (h) MnF$_2$, and (i) NiF$_2$. Error bars show the specific-heat uncertainties; vertical dotted lines mark the largest sampled values on each temperature grid.}
    \label{fig:cv_s3}
\end{figure}
\end{landscape}

\clearpage
\begin{landscape}
\begin{figure}[p]
    \centering
    \includegraphics[width=0.92\linewidth]{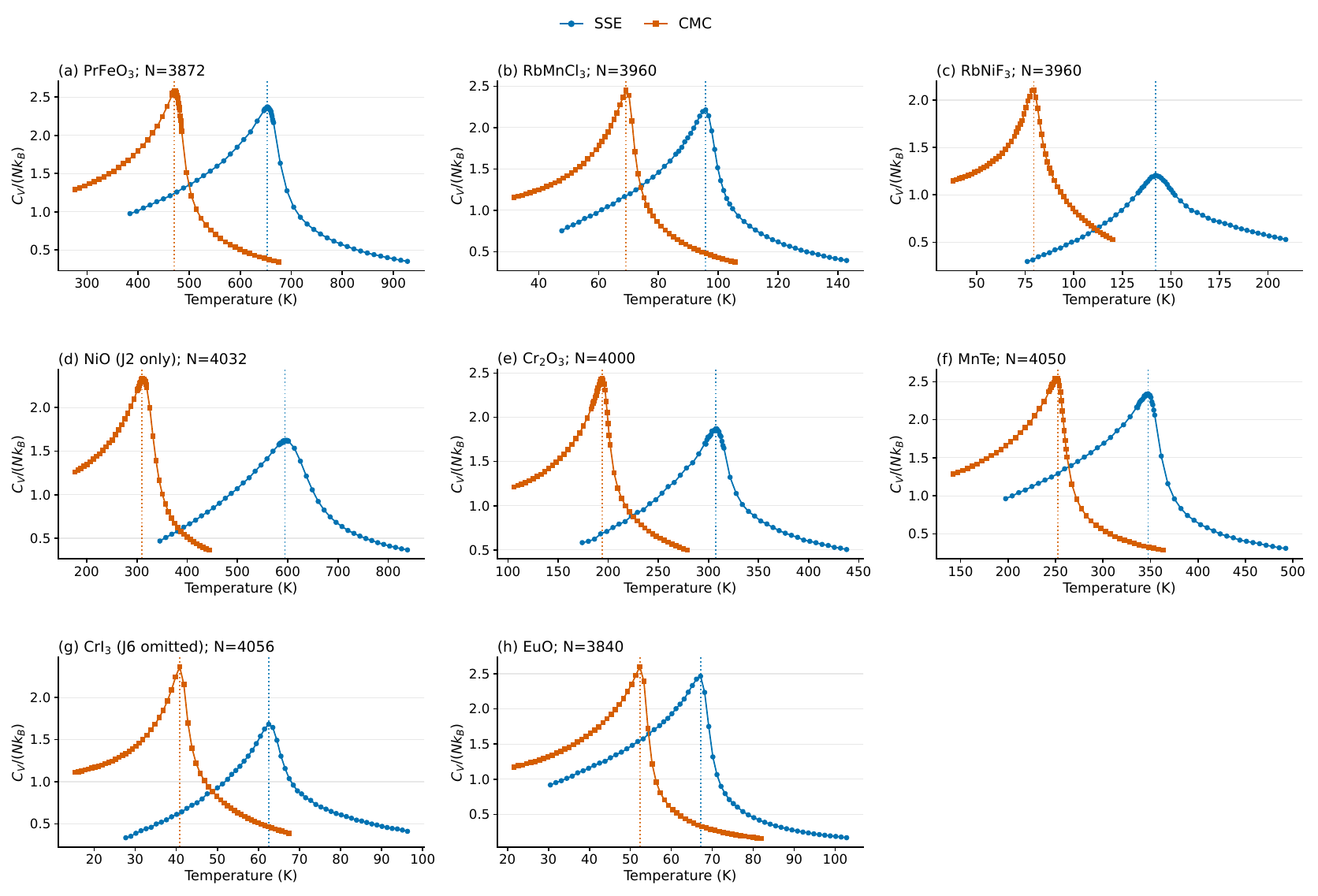}
    \caption{Specific heat per magnetic site from the approximately 4000-site SSE and CMC simulations for (a) PrFeO$_3$, (b) RbMnCl$_3$, (c) RbNiF$_3$, (d) NiO (J$_2$ only), (e) Cr$_2$O$_3$, (f) MnTe, (g) CrI$_3$ (J$_6$ omitted), and (h) EuO. Error bars show the specific-heat uncertainties; vertical dotted lines mark the largest sampled values on each temperature grid.}
    \label{fig:cv_s4}
\end{figure}
\end{landscape}
\clearpage
\section*{Magnetic energy differences and experimental ordering temperatures}
\begin{figure}[htbp]
    \centering
    \includegraphics[width=0.95\linewidth]{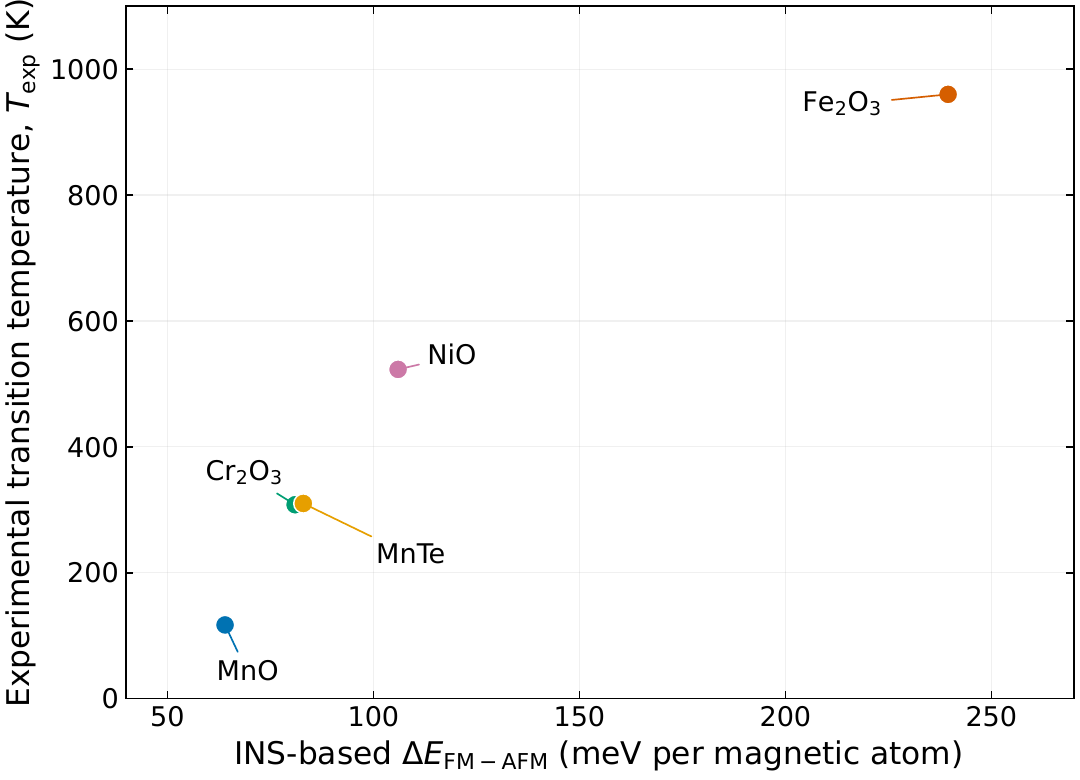}
    \caption{Experimental ordering temperature versus INS-based FM--AFM energy difference per magnetic atom for MnO, Cr$_2$O$_3$, MnTe, NiO, and Fe$_2$O$_3$. All five compounds are antiferromagnets, so the experimental temperatures are N\'eel temperatures. The temperatures are 117, 308, 310, 523, and 960~K, respectively. Cr$_2$O$_3$ and MnTe nearly overlap. The scatter plot illustrates the trend within this five-compound set; no universal proportionality is assumed and no fit is shown.}
    \label{fig:ins_energy_temperature}
\end{figure}

Figure~\ref{fig:ins_energy_temperature} compares the INS-based FM--AFM energy differences in Table I of the main text with experimental ordering temperatures for the five electronic-benchmark compounds. The temperatures are taken from the \texttt{T\_exp} column of \texttt{Compounds/T.dat} in the \texttt{Magnetic\_data} repository: \url{https://github.com/malaei/Magnetic_data/blob/main/INS_SW_MC/Compounds/T.dat}.

\clearpage
\providecommand{\bibcommenthead}{}
\bibliographystyle{sn-nature-linked}
\bibliography{literature}